\documentclass[twocolumn]{aastex62}
\usepackage{natbib}

\newcommand{\angstrom}{\textup{\angstrom}}

\graphicspath{{./}{figures_submit/}}

\shorttitle{A Candidate Type-II Quasar at $z=6.1292$}
\shortauthors{Onoue et al.}

\begin{document}

\title{Subaru High-z Exploration of Low-Luminosity Quasars (SHELLQs). XIV. A Candidate Type-II Quasar at $z=6.1292$}

\author[0000-0003-2984-6803]{Masafusa Onoue}
\email{onoue@mpia-hd.mpg.de}
\affiliation{Max-Planck-Institut f\"ur Astronomie, K\"onigstuhl 17, D-69117 Heidelberg, Germany}

\author[0000-0002-7402-5441]{Yoshiki Matsuoka}
\affiliation{Research Center for Space and Cosmic Evolution, Ehime University, 
Matsuyama, Ehime 790-8577, Japan}

\author[0000-0001-5493-6259]{Nobunari Kashikawa}
 \affil{Department of Astronomy, School of Science, The University of Tokyo,
 Tokyo 113-0033, Japan}
 \affil{Research Center for the Early Universe, The University of Tokyo, 7-3-1 Hongo, Bunkyo-ku, Tokyo 113-0033, Japan}
 
\author[0000-0002-0106-7755]{Michael A. Strauss}
\affiliation{Princeton University Observatory, Peyton Hall,
Princeton, NJ 08544, USA.}

\author[0000-0002-4923-3281]{Kazushi Iwasawa}
\affiliation{ICREA and Institut de Ci\`encies del Cosmos, Universitat de Barcelona, 
IEEC-UB, Mart\'i i Franqu\`es, 1, 08028 Barcelona, Spain}

\author[0000-0001-9452-0813]{Takuma Izumi}
\affiliation{National Astronomical Observatory of Japan, 
2-21-1, Osawa, Mitaka, Tokyo 181-8588, Japan}
 \affiliation{Department of Astronomical Science, 
Graduate University for Advanced Studies (SOKENDAI),
 2-21-1, Osawa, Mitaka, Tokyo 181-8588, Japan}

\author[0000-0002-7402-5441]{Tohru Nagao}
\affiliation{Research Center for Space and Cosmic Evolution, Ehime University, Matsuyama, Ehime 790-8577, Japan}

\author{Naoko Asami}
\affiliation{Seisa University, Hakone-machi, Kanagawa, 250-0631, Japan}

\author[0000-0001-7201-5066]{Seiji Fujimoto}
\affiliation{Cosmic Dawn Center (DAWN), Jagtvej 128, DK2200 Copenhagen N, Denmark}
\affiliation{Niels Bohr Institute, University of Copenhagen, Lyngbyvej 2, DK2100 Copenhagen {\O}, Denmark}

\author[0000-0002-6047-430X]{Yuichi Harikane}
\affiliation{Institute for Cosmic Ray Research, The University of Tokyo, 5-1-5 Kashiwanoha, Kashiwa, Chiba 277-8582, Japan}
\affiliation{Department of Physics and Astronomy, University Colege London, Gower Street, London WC1E 6BT, UK}

\author[0000-0002-0898-4038]{Takuya Hashimoto}
\affiliation{Tomonaga Center for the History of the Universe (TCHoU), Faculty of Pure and Applied Sciences, University of Tsukuba, Tsukuba, Ibaraki 305-8571, Japan}

\author[0000-0001-6186-8792]{Masatoshi Imanishi}
\affiliation{National Astronomical Observatory of Japan, 
2-21-1, Osawa, Mitaka, Tokyo 181-8588, Japan}
\affiliation{Department of Astronomical Science, 
Graduate University for Advanced Studies (SOKENDAI),
 2-21-1, Osawa, Mitaka, Tokyo 181-8588, Japan}

 \author[0000-0003-1700-5740]{Chien-Hsiu Lee}
 \affiliation{NSF’s National Optical-Infrared Astronomy Research Laboratory, 950 North Cherry Avenue, Tucson, AZ 85719, USA.}

\author{Takatoshi, Shibuya}
\affiliation{Kitami Institute of Technology, 165 Koen-cho, Kitami, Hokkaido 090-8507, Japan}

\author[0000-0002-3531-7863]{Yoshiki Toba}
\affiliation{Department of Astronomy, Kyoto University, Kitashirakawa-Oiwake-cho, Sakyo-ku, Kyoto 606-8502, Japan}
\affiliation{Academia Sinica Institute of Astronomy and Astrophysics, 11F of Astronomy-Mathematics Building, AS/NTU, No.1, Section 4, Roosevelt Road, Taipei 10617, Taiwan.}
\affiliation{Research Center for Space and Cosmic Evolution, Ehime University, 
Matsuyama, Ehime 790-8577, Japan}



\begin{abstract}
We present deep Keck/MOSFIRE near-infrared spectroscopy of a strong Ly$\alpha$ emitting source at $z=6.1292$, HSC J142331.71$-$001809.1, which was discovered by the SHELLQS program from imaging data of the Subaru Hyper Suprime-Cam (HSC) survey.
This source is one of five objects that show unresolved ($<230$ km s$^{-1}$) and prominent ($>10^{44}$ erg s$^{-1}$) Ly$\alpha$ emission lines at absolute 1450 \AA\ continuum magnitudes of $M_{1450}\sim-22$ mag.
Its rest-frame Ly$\alpha$ equivalent width (EW) is $370\pm30$ \AA.
In the 2 hour Keck/MOSFIRE spectrum in $Y$ band, the high-ionization C{\sc iv} $\lambda\lambda$1548,1550 doublet emission line was clearly detected with FWHM $=120^{+20}_{-20}$ km s$^{-1}$ and a total rest-frame EW of $37_{-5}^{+6}$ \AA.
We also report the detection of weak continuum emission, and the tentative detection of O{\sc iii]}$\lambda\lambda 1661,1666$ in the 4 hour $J$ band spectrum.
Judging from the UV magnitude, line widths, luminosities, and EWs of Ly$\alpha$ and C{\sc iv}, 
we suggest that this source is a reionization-era analog of classical type-II AGNs, although there is a possibility that it represents a new population of AGN/galaxy composite objects in the early universe.
We compare the properties of J1423$-$0018 to intermediate-redshift type-II AGNs and C{\sc iv} emitters seen in $z=6$--$7$ galaxy samples.
Further observations of other metal emission lines in the rest-frame UV or optical, and X-ray follow-up observations of the $z=6$--$7$ narrow-line quasars are needed for more robust diagnostics and to determine their nature.
\end{abstract}


\section{Introduction} \label{sec:Sec1}
The last two decades have yielded remarkable success in the discovery of unobscured quasars in the reionization epoch ($z>6$).
Thanks to the advent of wide-field surveys such as the Sloan Digital Sky Survey \citep[e.g.,][]{Jiang16} and Pan-Starrs1 \citep[e.g.,][]{Banados16}, more than 200 $z > 5.7$ quasars have been reported, with redshifts up to $z\sim7.6$ \citep{Banados18, Yang20, Wang21}.
Most of the luminous $z>6$ quasars (with bolometric luminosity $L_\mathrm{bol}>10^{46}$ erg s$^{-1}$) are powered by billion-solar-mass supermassive black holes (SMBHs), which is remarkable given the fact that the universe is less than one billion years old at $z>6$.
How such massive SMBHs could form at $z>6$ has been the focus of active discussion in the literature \citep[][for a recent review]{Inayoshi20}.

However, these optical (rest-frame UV) quasar surveys have not been sensitive to {\em obscured} quasars. 
Only a handful of candidate type-II active galactic nuclei (AGNs) are known at $z\gtrsim6$.  
They have been identified as either radio galaxies \citep{Saxena18} or companion X-ray sources of luminous unobscured quasars \citep{Connor19, Vito19}
\footnote{\citet{Vito21} recently reported no significant X-ray emission from \citet{Vito19}'s $z=6.515$ source with deeper follow-up observations with {\em Chandra}.}.
Luminous type-II AGNs whose bolometric luminosities are high enough to be classified as quasars have been identified at $z<4$ through observations at various wavelength ranges: optical \citep[e.g.,][]{Zakamska03, Alexandroff13, Yuan16}, infrared \citep[e.g.,][]{Stern12, Lacy15, Glikman18}, X-ray \citep[e.g.,][]{Stern02, Szokoly04}, and radio \citep[e.g.,][]{vanBreugel99, DeBreuck00}.
The intrinsic luminosities and degree of obscuration depend strongly on the selection criteria, making it challenging to compare results from different samples.  
Yet, at $z<0.8$ \citet{Reyes08} suggest that type-II quasars are as abundant as type-I quasars at fixed [O{\sc iii}] $\lambda5007$ luminosity ($L_\mathrm{[OIII]}>10^{8.3}L_\odot$).
Sensitive X-ray observations also showed that the obscured ($N_\mathrm{H} > 10^{23}$ cm$^{-2}$) fraction is $>60$\%\ at $3<z<6$, increasing from lower redshift \citep{Ueda03, Ueda14, Vito18}.
Those high-redshift obscured type-II quasars and AGNs are useful probes of the metal enrichment of host galaxies, because their narrow emission lines trace host-scale ionized gas, unlike broad emission lines from the nuclear region.
Radio galaxy observations have found no significant redshift evolution of the narrow-line-region metallicity up to $z\sim5$ \citep{Nagao06b, MatsuokaK09, MatsuokaK11, Maiolino19}.

Obscuration of AGNs can occur at different scales within the host galaxies.
The standard AGN unification models explain the two types of AGNs as due to different viewing angles to the central accretion disks \citep[e.g.,][]{Antonucci93, Urry95}.
In this framework, type-II AGNs are observed when the observers' lines of sight are obscured by optically thick dusty material that blocks the nuclear emission at $\lesssim10$ pc from the central SMBHs.
An alternative model suggests that the obscured AGNs constitute a transitional phase during a gas-rich galaxy major merger \citep[e.g.,][]{Hopkins06}; obscured AGNs appear when SMBHs first ignite, surrounded by dust in the host galaxy generated by a merger-induced starburst.
In these models, one predicts a population of dust-reddened broad-line quasars before the strong radiation pressure from the AGN completely expels the surrounding dust.
This modestly obscured population, so-called red quasars and dust-obscured galaxies, have been observed over a broad range of redshift \citep[e.g.,][]{Richards03, Urrutia08, Glikman12, Assef13, Ross15, Toba18, Kato20}.
Theoretical studies have recently suggested that the initial intense growth of SMBHs at high redshift is accompanied by high column-density gas and dust in the host galaxies, and such host-scale gas and dust can significantly contribute to the observed optical depth \citep{Trebitsch19, Davies19, Ni20}.
Thus there is particular interest in identifying red and obscured AGN at high redshift, because
those type-II objects may provide clues to the initial growth of SMBHs as well as dust and chemical enrichment in the host galaxies.

\subsection{Narrow-Line Population in the $z=6$--$7$ Low-Luminosity Quasar Sample} \label{sec:Sec1_1}
A $1000$ deg$^2$-class optical survey with the Hyper Suprime-Cam \citep{Miyazaki18} mounted on the 8.2m Subaru telescope  \citep{SSPpaper} has enabled the deepest investigation to date of the faint end of the quasar luminosity function at high redshift.
With this survey, the Subaru High-$z$ Exploration Low-Luminosity Quasar (SHELLQs) project has found 93 quasars at $5.7\leq z\leq7.1$ in the low-luminosity range
(down to bolometric luminosity $L_\mathrm{bol}\sim10^{45}$ erg s$^{-1}$, or 1450 \AA\ absolute magnitude $M_{1450}\sim -22$) 
\citep{Matsuoka16, Matsuoka18a,Matsuoka18b,Matsuoka19a, Matsuoka19b}.
Those objects were selected with a standard color cut that makes use of red $i-z$ and $z-y$ colors of $z\sim6$ and $z\sim7$ quasars, respectively.
In the SHELLQs project, quasar candidates are required to be point sources in their HSC images (typical seeing of $0.7$ arcsecond in the $z$ and $y$ bands).
Their UV magnitudes were derived by extrapolating the observed optical continua (rest-frame $\lambda_\mathrm{rest}\approx1220$--$1350$\AA) to rest-frame 1450\AA, assuming the fiducial power-law slope \citep[e.g.,][]{VB01} of $\alpha_\lambda=-1.5$   ($F_\lambda\propto\lambda^{\alpha_\lambda}$).
The Ly$\alpha$ properties were measured by subtracting the underlying continuum flux that is estimated by pixels redward of Ly$\alpha$. 
Please refer to \citet{Matsuoka19b} for more details on the sample selection and discovery observations.

The luminosity function of $z=6$ quasars overlaps that of dropout galaxies at $M_\mathrm{1450}=-23$ mag \citep{Ono18, Matsuoka18c}; therefore we cannot assume that the observed rest-UV continua and emission lines of the faintest SHELLQs objects are purely from the nuclear regions.
The host stellar light contamination to the rest-UV quasar continuum at $z>4$ has recently been discussed both theoretically \citep{Trebitsch20} and observationally \citep{Adams20, Bowler21}.

\begin{figure*}[htbp]
\centering
 \includegraphics[width=\linewidth]{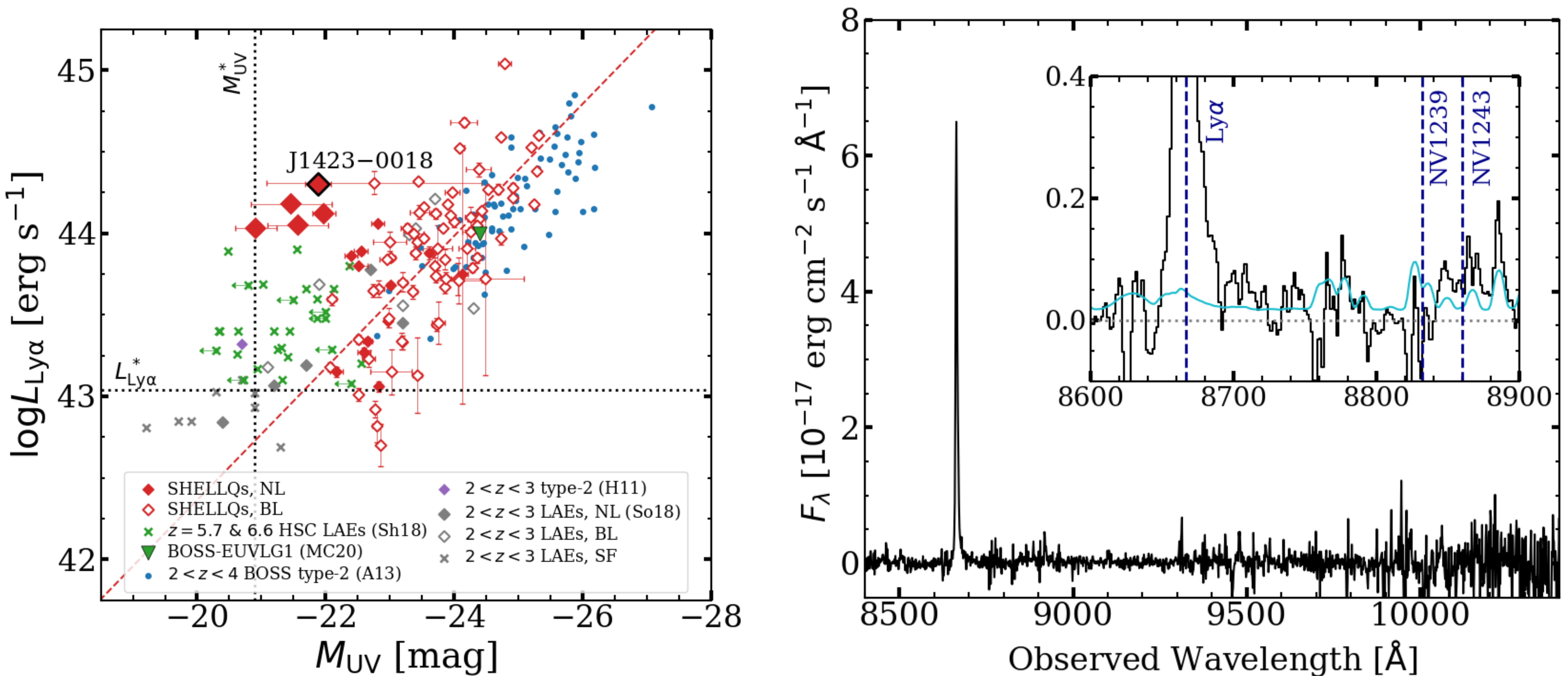}
\caption{
({\it a}) The absolute UV magnitude - Ly$\alpha$ luminosity plane of quasars and LAEs.
The $z=6$--$7$ narrow-line and broad-line SHELLQs quasars  \citep{Matsuoka16, Matsuoka18a, Matsuoka18b, Matsuoka19a, Matsuoka19b} are shown as red filled and open diamonds, respectively.
The five extreme narrow-line objects with $M_\mathrm{UV}>-22$ and $\log{Ly\alpha}>44$ are highlighted with large symbols.
The red dashed line shows a linear fit to the broad-line quasars.
For comparison, spectroscopically confirmed HSC LAEs at $z=5.7$ and $6.6$  \citep{Shibuya18} are shown as green crosses.
A luminous star-forming galaxy at $z=2.5$ (BOSS-EUVLG1; \citealt{MC20}) at $M_\mathrm{UV}=-24.40\pm0.05$ mag and $\log{L_\mathrm{Ly\alpha}=44.0\pm0.1}$ erg s$^{-1}$ is shown as a green triangle.
Type-II quasar candidates at $2<z<4$ from the SDSS BOSS survey  \citep[][their Class A objects]{Alexandroff13} are indicated with blue dots.
The composite spectrum of narrow-line AGNs at $2<z<3$ from \citet{Hainline11} is shown as a purple diamond.
The $2<z<3$ LAEs from \citet{Sobral18} are shown in grey, for which three different symbols indicate different classifications: narrow-line AGNs (FWHM$_\mathrm{Ly\alpha}$ $<1000$ km s$^{-1}$; filled diamonds), broad-line AGNs (FWHM$_\mathrm{Ly\alpha}$ $\geq1000$ km s$^{-1}$; open diamonds), and galaxies (crosses).
The Ly$\alpha$ luminosity of J1423$-$0018 is significantly larger than those of the broad-line SHELLQs quasars of similar UV continuum magnitude, as well as those of known LAEs.
Horizontal and vertical dashed lines show the knee of the Ly$\alpha$ luminosity function at $z=5.7$ \citep[$L_\mathrm{Ly\alpha}^*=1.1\times10^{43}$ erg s$^{-1}$;][]{Konno18}, and the
UV luminosity function at $z\sim6$ \citep[$M_\mathrm{UV}^*=-20.9$ mag;][]{Ono18}.
({\it b})
The optical spectrum of J1423$-$0018 taken by Subaru/FOCAS \citep{Matsuoka18a}.
The inset shows the Ly$\alpha$ wing and the expected location of N{\sc v} $\lambda\lambda1239, 1243$.
The spectrum in the inset is smoothed with a Gaussian kernel of $\sigma=1$ pixel.
The error spectrum is shown in cyan.
} \label{fig:fig1}
\end{figure*}

Intriguingly, 16  of the spectroscopically confirmed SHELLQs quasars ($\sim 20\%$ of the sample) show narrow Ly$\alpha$ emission with FWHM $<500$ km s$^{-1}$ \citep{Matsuoka18a, Matsuoka19b}, while typical broad emission lines of type-I quasars have FWHM $=$ several $1000$ km s$^{-1}$.  
We tentatively refer to this population as narrow-line quasars, as their Ly$\alpha$ luminosities of $10^{43-44}$ erg s$^{-1}$ are in the range of AGNs at lower redshifts \citep{Sobral18, Spinoso20}, and are comparable to those of the $2<z<4$ type-II quasar candidates of \citet{Alexandroff13}.
Another possible AGN signature is seen in a weak P-Cygni-like profile of the high-ionization N{\sc v} $\lambda1240$ line (ionization potential: 77.4 eV) tentatively detected in the composite spectrum of the narrow-line objects \citep{Matsuoka18a, Matsuoka19b}.
The low signal-to-noise ratio of the spectra precludes identifying the presence of interstellar absorption in the rest-frame UV continuum.
Figure~\ref{fig:fig1}a shows that the narrow-line objects have some of the lowest UV continuum luminosities in the SHELLQs sample.  
However, their Ly$\alpha$ luminosities are significantly higher than those of the broad-line SHELLQs quasars with similar UV magnitudes, suggesting that they comprise a distinct population.
Their Ly$\alpha$ brightness is comparable to, or even larger than those of $z=5.7$ and $z=6.6$ Lyman alpha emitters (LAEs) identified in the deep and ultradeep layers of the HSC survey \citep{Shibuya18}, as well as luminous $2<z<3$ LAEs presented in \citet{Sobral18}.
Five sources stand out as being very faint in the UV continuum ($M_\mathrm{UV}>-22$ mag) and extremely bright in Ly$\alpha$ ($L_\mathrm{Ly\alpha}>10^{44}$ erg s$^{-1}$) \citep{Matsuoka18a, Matsuoka19b}.
In Figure~\ref{fig:fig1}a we highlight those five objects with large symbols.
Those features cause us to speculate that the SHELLQs narrow-line population represents the obscured counterpart to the broad-line quasars at the reionization epoch.  This population is so faint optically that it could not be probed with previous shallower surveys.

\begin{figure*}[htbp]
\centering
 \includegraphics[width=\linewidth]{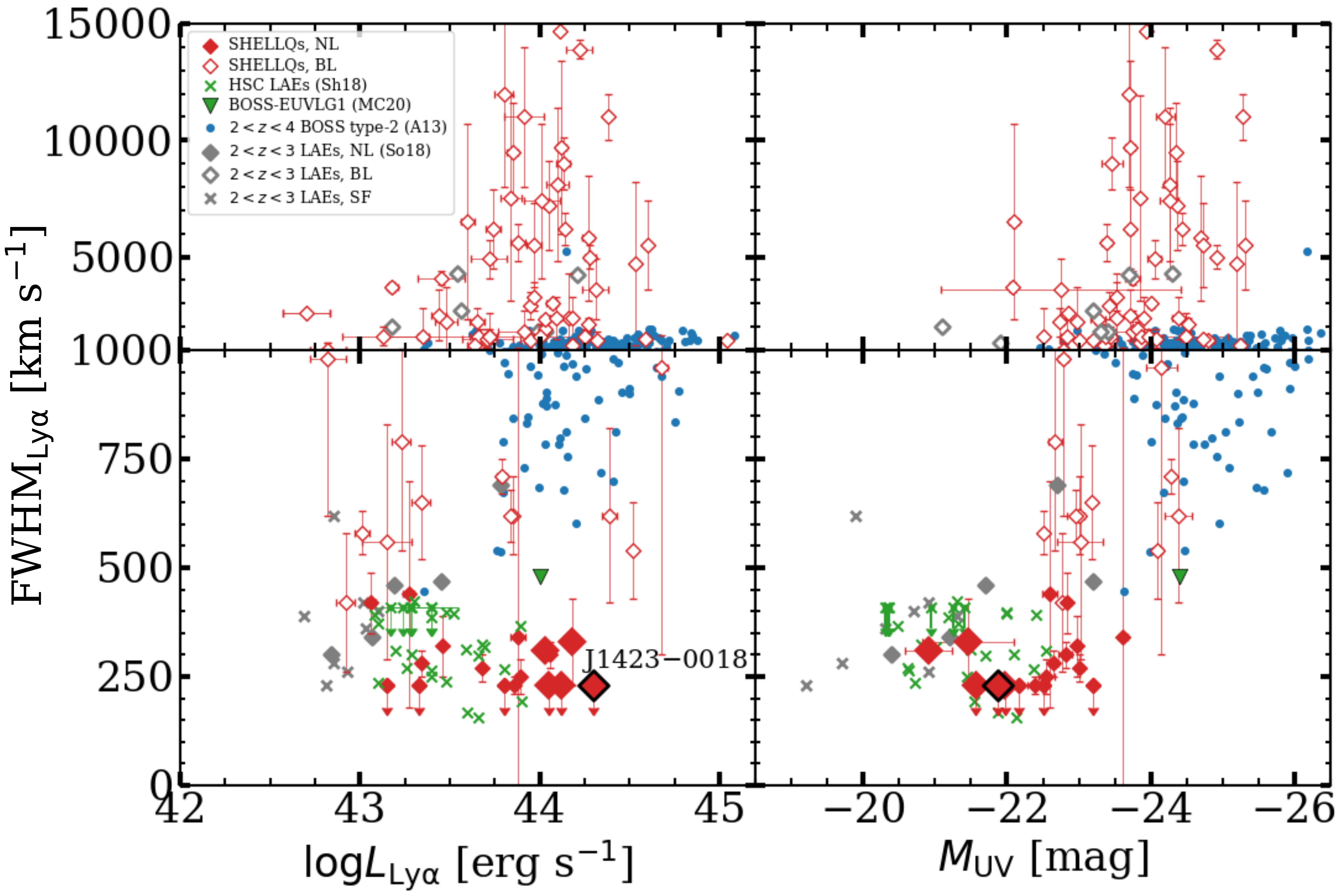}
\caption{
Ly$\alpha$ FWHM versus Ly$\alpha$ luminosities ({\it left}) and UV magnitudes ({\it right})  for the objects shown in  Figure~\ref{fig:fig1}a.  
The symbol types are the same as in that figure. 
Note that the y-axis scale is different in the lower and upper panels.  
The five extreme SHELLQs NL objects, including J1423$-$0018, lie in the most Ly$\alpha$ luminous ($\log{L_\mathrm{Ly\alpha}}>44$ [erg s$^{-1}$]) and UV continuum faintest ($M_\mathrm{UV}>-22$ mag) ranges.
} \label{fig:fig2}
\end{figure*}

In this paper, we present Keck/MOSFIRE follow-up observations of HSC J142331.71-001809.1 (hereafter J1423$-$0018) at $z=6.1292$.
This object was discovered in \citet{Matsuoka18a}.
J1423$-$0018 is one of the five extreme narrow-line SHELLQs quasars, and it was chosen for near-infrared spectroscopy because it has the highest Ly$\alpha$ luminosity ($L_\mathrm{Ly\alpha}=10^{44.30\pm0.01}$ erg s$^{-1}$) of all the narrow-line objects discovered in SHELLQs. 
The rest-frame equivalent width of Ly$\alpha$ is EW$_\mathrm{rest}$(Ly$\alpha$)$=370\pm30$ \AA.
Figure~\ref{fig:fig1}b shows its discovery spectrum in the optical. 
It is one of the faintest SHELLQs sources in the continuum  ($M_\mathrm{1450}=-21.88\pm0.20$ mag); its Ly$\alpha$ redshift is $z=6.13$. 
Note that N{\sc v} $\lambda\lambda1239,1243$ is not detected, but there are strong OH sky emission lines at the wavelengths of the expected line centers.
This object shows a remarkably large Ly$\alpha$ rest-frame equivalent width of EW$_\mathrm{rest}$(Ly$\alpha$)$=370\pm30$ \AA, and the highest Ly$\alpha$ luminosity among the narrow-line population, with $L_\mathrm{Ly\alpha}= 10^{44.30\pm0.01}$ erg s$^{-1}$.
The Ly$\alpha$ line is not resolved (FWHM $<230$ km s$^{-1}$) in the discovery spectrum (Figure~\ref{fig:fig1}b).
Figure~\ref{fig:fig2} plots Ly$\alpha$ FWHM as a function of Ly$\alpha$ and UV luminosities for J1423$-$0018 and the objects shown in Figure~\ref{fig:fig1}a.

Throughout this paper, all magnitudes quoted are on the AB system.
We adopt a standard $\Lambda$CDM cosmology with $H_0=70$ km s$^{-1}$ Mpc$^{-1}$, $\Omega_m=0.3$, and $\Omega_\Lambda=0.7$.

\section{Keck/MOSFIRE Spectroscopy} \label{sec:obs}
We carried out near-infrared spectroscopic observations of J1423$-$0018 with MOSFIRE \citep{McLean12} on the Keck I telescope on
the first half of the nights of 2019 June 6 and 7 (program ID: S19A-102\footnote{Subaru-Keck exchange program}).
J1423$-$0018 was observed in the $Y$ and $J$ bands together with filler sources, covering the wavelength ranges of $9716$--$11250$ \AA\ and $11530$--$13520$ \AA, respectively.
Those filler sources included a $z\sim1$ radio galaxy and two $z\sim4$ quasar candidates newly identified in the HSC survey.
A more detailed description and analysis of those objects will be presented elsewhere. The mask also included a point source  for flux calibration of the science targets, as described below. 
Those five targets were aligned to custom slitmasks with bright reference stars in the target field.
We selected the $0\farcs7$-wide slit to perform our spectroscopy, giving spectral resolutions of $R=3388$ in the $Y$ band and $R=3318$ in the $J$ band.
Individual exposure times were 180 seconds in $Y$ and 120 seconds in $J$, respectively.
The total exposure times were $2.0$ hours in $Y$ and $3.8$ hours in $J$.
During the exposures, we employed a four point (ABA$^\prime$B$^\prime$) dither pattern to maximize the quality of the faint source spectra.
The nodding amplitudes of the AB and A$^\prime$B$^\prime$ pairs were 2.7 arcseconds, or 15 pixels, with a relative offset of 0.3 arcseconds between the two pairs. 
During the observations, the seeing ranged from $0\farcs4$ to $0\farcs9$ and the targets were observed at airmass $1.1$--$1.5$.

The raw data were reduced with the MOSFIRE Data Reduction Pipeline in the standard way.
One-dimensional spectra were extracted with a boxcar aperture twelve pixels ($2\farcs2$) wide, roughly twice the seeing.  
This wide aperture was motivated by the spatial extent of C{\sc iv} $\lambda\lambda$ 1548,1550 and O{\sc iii]} $\lambda1666$, and 
the apparent extended O{\sc iii]}$\lambda1661$ signal visible in the two-dimensional $J$ band spectrum (Section~\ref{sec:results}).
The central pixel of the object trace was determined from the spatial profile of the C{\sc iv} emission lines.
The positive signal that we interpret as O{\sc iii]} $\lambda1666$ (Section~\ref{sec:results}) is centered on the extraction aperture.
Telluric absorption was corrected with long-slit observations of A0-type stars.  
The point source observed in the field ($\alpha,\delta = 14$:23:47.81,$-$00:17:52.2) has NIR photometry from the VIKING DR3 (\citealt{Edge13}; $Y_\mathrm{AB}=17.966 \pm 0.009$, $J_\mathrm{AB}=17.811 \pm 0.009$); we used this for spectrophotometric flux calibration.  
The $Y$ band exposures taken during the two nights were all stacked before one-dimensional extraction to efficiently remove cosmic-ray hits on the detectors, while the $J$ band exposures were stacked separately for each night before extraction. 
The final $J$ band spectrum is an inverse-variance weighted mean of the spectra from the two nights.

\section{Results} \label{sec:results}
The MOSFIRE spectrum covers various UV emission lines at rest-frame $1370$--$1895$ \AA.
We measured the emission line fluxes (or upper limits) of Si{\sc iv} $\lambda\lambda$ 1393, 1402, N{\sc iv} $\lambda\lambda$ 1483, 1487, C{\sc iv} $\lambda\lambda1548$, $1550$ in the $Y$ band, and He{\sc ii} $\lambda1640$, O{\sc iii]} $\lambda\lambda1661$, $1666$, and Si{\sc iii} $\lambda\lambda1883$, $1892$ in the $J$ band.

\begin{figure*}[htbp]
\centering
 \includegraphics[width=0.95\linewidth]{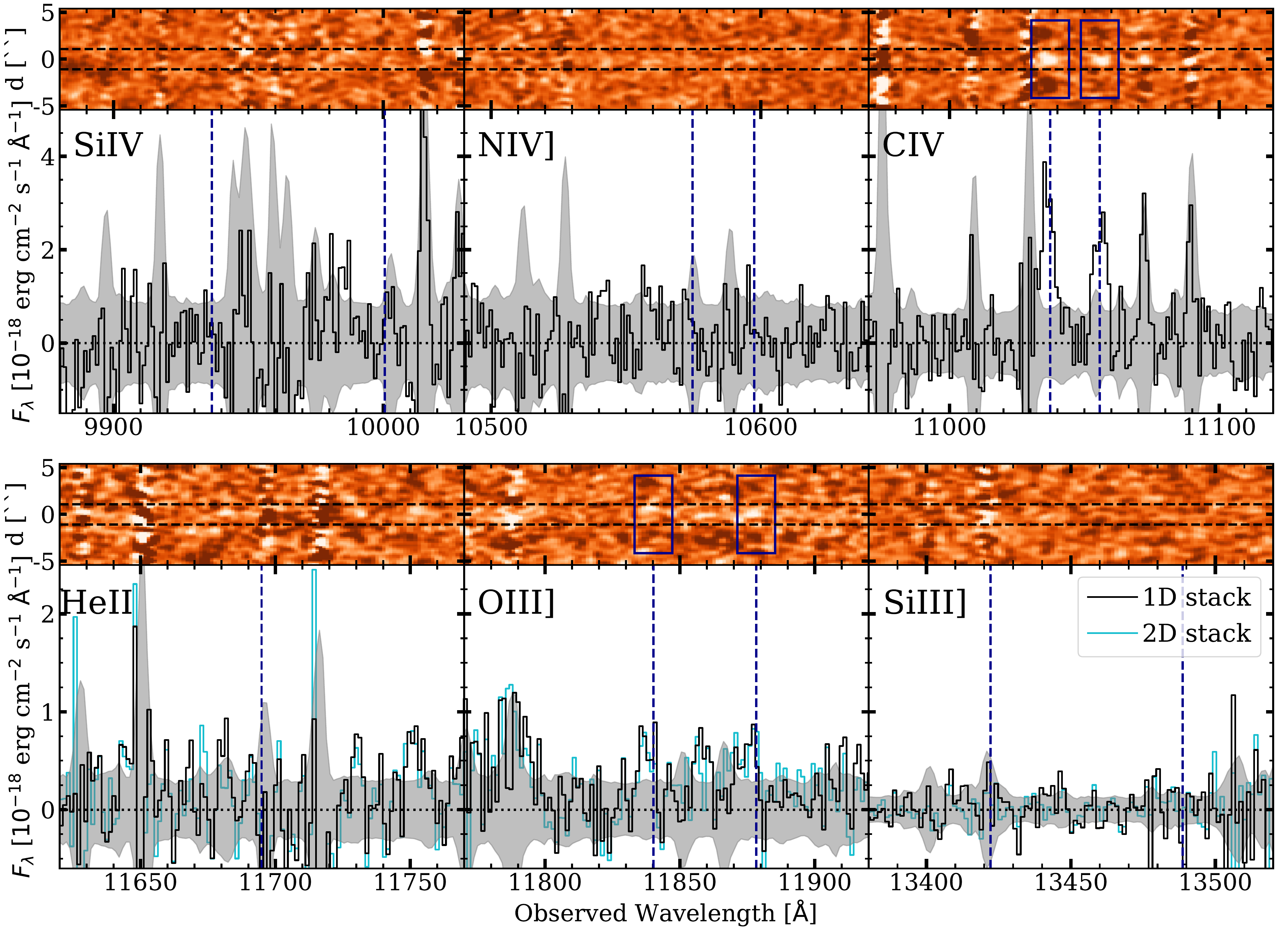}
\caption{
The Keck/MOSFIRE spectrum of J1423$-$0018 around the rest-frame UV emission lines (Si{\sc iv} $\lambda\lambda$ 1393, 1402, N{\sc iv]} $\lambda\lambda$ 1483, 1487, and C{\sc iv} $\lambda\lambda$ 1548, 1550 in the first row, and He{\sc ii} $\lambda$ 1640, O{\sc iii]} $\lambda\lambda$ 1661, 1666, and Si{\sc iii]} $\lambda\lambda$ 1883, 1892 in the second row). 
Both two and one dimensional spectra are shown for each line region.  
The two-dimensional spectrum is not corrected for telluric absorption, nor has been relative/absolute flux calibrated.  
It has been smoothed with a Gaussian kernel with $\sigma=1$ pixel, where positive/negative signals are shown in white/black.
The extraction aperture in the two-dimensional spectra is indicated with black dashed lines.
The $\pm1\sigma$ flux error is shown as the gray shading in the one-dimensional spectra.
The line centers expected from the measured C{\sc iv} redshift are marked in the one-dimensional spectra with dark-blue dashed lines.
For the $J$ band, we also show the one-dimensional spectrum obtained by stacking all the two-dimensional frames taken on the two observing dates before extraction (cyan).
} \label{fig:fig3} 
\end{figure*}

The top two panels of Figure~\ref{fig:fig3} showcase the MOSFIRE spectrum at the wavelengths of those emission lines.  
The  C{\sc iv} $\lambda \lambda$1548, 1550 doublet, which is usually observed as a blended single line in a broad-line quasar, was clearly detected as two distinct lines. 
Note in particular both the positive and negative signals from each of the lines of the doublet in the two-dimensional spectrum. 
The detection significance of the two lines in the one-dimensional spectrum is $6.6$ and $4.5$ sigma, respectively, when the continuum-subtracted signals are integrated over $\pm 2$ FWHM from the line centers.

\begin{figure*}[hbtp]
\centering
 \includegraphics[width=\linewidth]{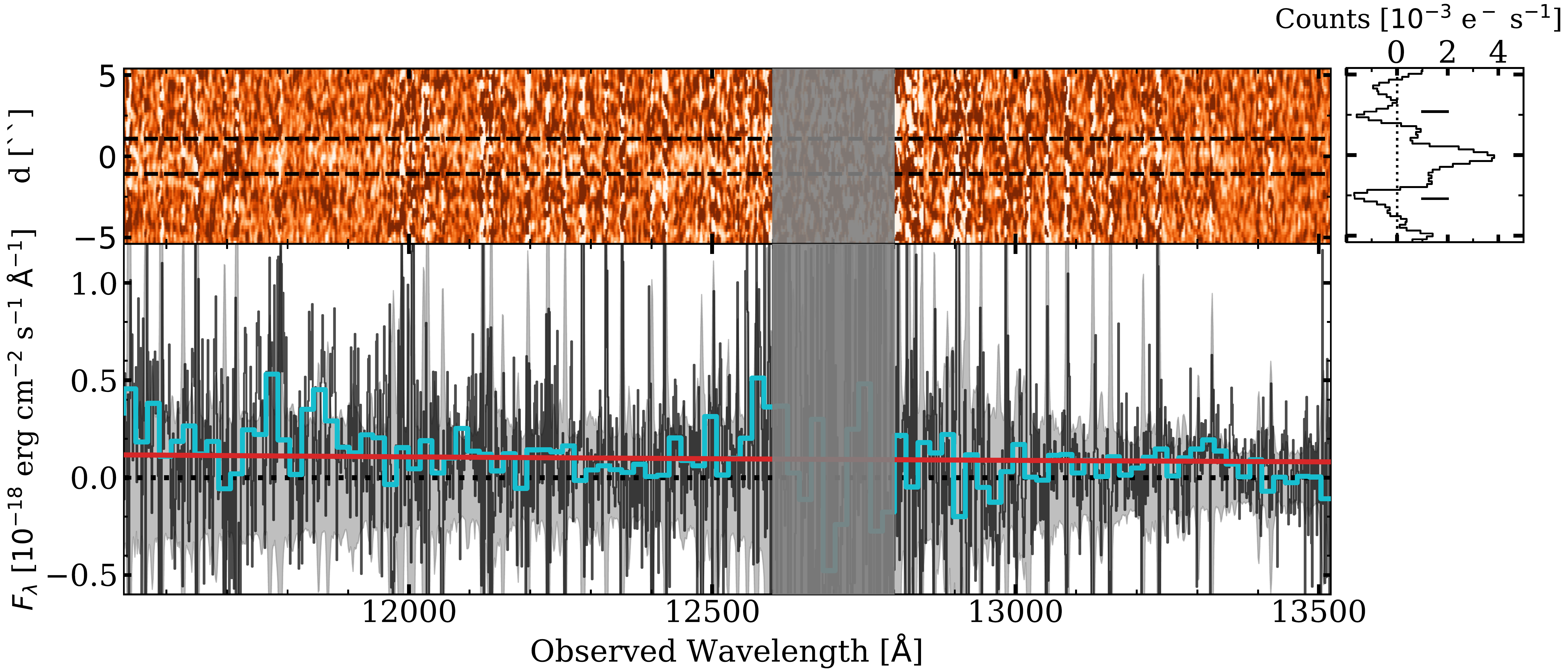}
\caption{
The full two- ({\it top}) and one-dimensional ({\it bottom}) $J$ band spectrum.
The two-dimensional spectrum is smoothed with a $\sigma=1$ pixel Gaussian kernel.
In the bottom panel, the cyan shows the 15-pixel binned spectrum ($19.5$\AA per bin).
The best-fit power-law continuum model with $\beta=-2.3$, for which $\lambda_\mathrm{obs}=12600$--$12800$ \AA\ is masked, is shown in red.
The masked wavelength range is shown in gray.
The right panel is the inverse-variance-weighted mean spatial profile of the smoothed spectrum, in which the averaged continuum flux is detected within the extraction aperture.
The expected positions of the negative traces are marked with tick marks.
} \label{fig:fig4} 
\end{figure*}

\begin{deluxetable*}{lCCCCC}[htb]
\tablecaption{Emission line properties of J1423$-$0018 \label{tab:emission_line}}
\tablecolumns{6}
\tablenum{1}
\tablewidth{0pt}
\tablehead{
\colhead{Line} &
\colhead{$\lambda_\mathrm{rest}$ } &
\colhead{$\lambda_\mathrm{obs} $ } &
\colhead{$F_\mathrm{line}$ } &
\colhead{EW$_\mathrm{rest}$}  &
\colhead{$\Delta v_\mathrm{Ly\alpha}$}  \\
\colhead{} &
\colhead{($\mathrm{\AA}$)} &
\colhead{($\mathrm{\AA}$)} &
\colhead{($10^{-17}$ erg cm$^{-2}$ s$^{-1}$)} &
\colhead{($\mathrm{\AA}$)} &
\colhead{(km s$^{-1}$)} 
}
\startdata
 Ly$\alpha$   & 1215.67  & 8667.7  & 47.60\pm0.01 & 370\pm30 &  \cdots  \\
 N{\sc v}     & 1238.82  & sky & \cdots & \cdots & \cdots   \\
 $\cdots$     & 1242.80  & sky  &  \cdots & \cdots & \cdots   \\
 Si{\sc iv}   & 1393.76  & \cdots & < 0.73 & < 6.3  & \cdots \\
 $\cdots$     & 1402.77  & \cdots &  < 0.91 & < 8.3  & \cdots \\
 N{\sc iv}    & 1483.3   & sky    & \cdots & \cdots & \cdots \\
 $\cdots$     & 1486.5   & \cdots & < 0.74 & < 7.5 & \cdots \\
 C{\sc iv}    & 1548.19  & 11037.4_{-0.3}^{+0.4} &  1.9_{-0.2}^{+0.3} & 21_{-3}^{+4}  & -32^{+8}_{-7} \\
 $\cdots$     & 1550.77  & 11055.8_{-0.3}^{+0.4} &  1.5_{-0.3}^{+0.3} & 16_{-3}^{+3}  &  \\
 He{\sc ii}   & 1640.42  & sky & \cdots &  \cdots  & \cdots \\
 O{\sc iii]}  & 1660.81  & 11838.0_{-0.6}^{+0.6} &   0.42_{-0.11}^{+0.13} & 5.3_{-1.5}^{+1.7}  & -90^{+14}_{-16} \\
 $\cdots$     & 1666.15  & 11876.1_{-0.6}^{+0.6} &  0.41_{-0.11}^{+0.13} & 5.3_{-1.4}^{+1.6}  &  \\
 Si{\sc iii]} & 1882.71  & sky  &  \cdots  &  \cdots  & \cdots \\
 $\cdots$     & 1892.03  & \cdots  & < -0.02 & < -0.4 & \cdots
\enddata
\tablecomments{
The emission line fluxes and rest-frame equivalent widths were measured after subtracting the continuum emission, which was estimated by fitting a $\beta=-2.3$ continuum to the $J$ band spectrum.
The $2\sigma$ upper limits of the line fluxes and rest-frame equivalent widths are provided for undetected lines, except for those overlapping with strong OH sky emission lines.
The Ly$\alpha$ line flux and rest-frame EW are taken from \citet{Matsuoka18a}.
For the velocity offsets with respect to Ly$\alpha$ ($\Delta v_\mathrm{Ly\alpha}$), negative values indicate blueshifts of the metal lines.
The quoted line flux of the Si{\sc iii]} doublet, which falls at the red edge of the $J$ band spectrum, is negative because the observed flux is smaller than the power-law continuum model.
}
\end{deluxetable*}

Figure~\ref{fig:fig4} shows that the trace of a weak continuum is visible in the $J$ band.
The robustness of the continuum detection is supported by the mean spatial profile of the Gaussian-smoothed ($\sigma=1$ pixel) two-dimensional spectrum.
The extracted one-dimensional spectrum (with 15-pixel binning) is clearly positive over the $J$ band.
The spectroscopic magnitude of this continuum component is $J=24.71\pm0.24$ mag\footnote{The $J$-band spectrum was convolved with the $J$-band transmission curve of UKIRT/WFCAM.}.  
We masked wavelengths between 1.26 and 1.28 microns in determining the continuum level due to the presence of strong lines of O$_2$ in the atmosphere. 
There is no $J$-band photometry available for this source.
Fitting the continuum of J1423$-$0018 to a power law, we estimated the UV slope to be $\beta=-2.3\pm1.2$, where $\beta$ is defined as $F_\lambda\propto\lambda^\beta$, based on the HSC-$y$ and the spectroscopic $J$-band magnitudes.
This continuum slope is bluer than that of the $z\sim2$ narrow-line quasars in \citet[][$\beta=-0.5$]{Hainline11}, but is consistent with those of UV dropout galaxies \citep[$-2.5<\beta<-1.5$;][]{Bouwens20}.
We fit the $\beta=-2.3$ power-law continuum model to the unbinned $J$-band spectrum to estimate the absolute $1450$ \AA\ magnitude.  We find $M_{1450}=-22.15_{-0.07}^{+0.10}$ mag, which is slightly brighter than the value of $M_{1450}=-21.88\pm0.20$ mag reported by \citet{Matsuoka18a}.

The results of our line measurements are reported in Table~\ref{tab:emission_line}.
We assumed that C{\sc iv}$\lambda1548$ and C{\sc iv}$\lambda1550$  originate from the same ionized gas, and thus fit them to the same line widths and redshifts. 
The errors quoted in Table~\ref{tab:emission_line}
were derived by a Monte Carlo simulation of 1000 mock spectra, in which random noise was added to the original spectrum given the estimated flux error at each pixel.  
Therefore, the uncertainty of our continuum measurement is taken into account in the error budgets of the emission line properties.
The $1\sigma$ uncertainties were determined from the 16$^{\rm th}$ and 84$^{\rm th}$ percentiles.
We measured a C{\sc iv} redshift of $z_\mathrm{CIV}=6.1292\pm0.0002$.
There is a modest C{\sc iv} velocity blueshift of $\Delta v_\mathrm{CIV - Ly\alpha}=-30\pm9$ km s$^{-1}$ with respect to Ly$\alpha$.
This C{\sc iv} redshift is our best estimate at this point of the systemic redshift of J1423$-$0018, although C{\sc iv} is a resonant line that is possibly red/blueshifted from nebular emission lines.
The velocity offset between Ly$\alpha$ and C{\sc iv} may be attributed to the fact that kinematics of the gas is different between neutral and highly ionized gas \citep{Steidel10}.
The line width after correcting for instrumental broadening is FWHM $=120_{-20}^{+20}$ km s$^{-1}$.  
A fit forced to the instrumental resolution of FWHM$=88$ km s$^{-1}$ gave a significantly worse $\chi^2$.  
Thus we are confident that the C{\sc iv} doublet line is resolved.
Both lines of the doublet are strong, with the flux ratio of C{\sc iv}$\lambda1548$ $/$ C{\sc iv} $\lambda1550$ $= 1.3$.
Their rest-frame equivalent widths are EW$_\mathrm{rest}=21_{-3}^{+4}$ \AA\ and $16_{-3}^{+4}$ \AA, respectively; therefore the total equivalent width is EW$_\mathrm{rest}=37_{-5}^{+6}$ \AA.

We searched for other emission lines and found weak signals at the wavelengths where the O{\sc iii]} doublet lines  ($\lambda\lambda1661$, $1666$) are expected based on the C{\sc iv} redshift.
There is apparent O{\sc iii]} $\lambda1666$ emission at $\lambda_\mathrm{obs}\sim11875$ \AA\ in the stacked spectrum.
This flux excess is clear in the stacked spectra of each of the  observing dates.
Flux excess is also seen at $\lambda_\mathrm{obs}\sim11840$ \AA\ where O{\sc iii]} $\lambda1661$ is expected.
This signal is slightly offset from the aperture center (Figure~\ref{fig:fig3}) determined by C{\sc iv}, but it is included in the 2.2 arcsecond aperture that we employed to measure all the detected signal from C{\sc iv} and O{\sc iii]} $\lambda1666$.
The detection significance of the O{\sc iii]} doublet emission lines is modest, at $3.3$ and $3.0$ sigma for O{\sc iii]} $\lambda1661$ and O{\sc iii]} $\lambda1666$, respectively.

In addition, there is a positive flux excess {\em between} the two O{\sc iii]} peaks at $\lambda_\mathrm{rest}\sim1663$ \AA\, with a significance of  $1$--$2$ sigma per pixel in the 1D spectrum.  
This feature  remains when all the two-dimensional frames taken on the two observing dates were stacked before extraction (cyan line in Figure~\ref{fig:fig3}).
Since its peak flux density is comparable to the two O{\sc iii]} lines and its nature is unclear, we conclude that the excess at $\lambda_\mathrm{rest}\sim1666$ \AA\ and especially at 1661\AA\ cannot be conclusively attributed to O{\sc iii]} emission.

Despite this uncertainty, we measured the O{\sc iii]} profiles while ignoring the third peak at $\lambda_\mathrm{rest}\sim1663$ \AA.  As with our C{\sc iv} fit, we fit 
 each of the O{\sc iii]} $\lambda1661$ and O{\sc iii]} $\lambda1666$ lines with a single Gaussian profile with a common velocity offset with respect to Ly$\alpha$.
The line width was fixed to that of C{\sc iv} convolved with the $J$-band resolution. 
The O{\sc iii]}  velocity offset is $\Delta v_\mathrm{OIII]-Ly\alpha}=-90^{+14}_{-16}$ km s$^{-1}$ ($z_\mathrm{OIII]}=6.1279\pm0.0004$), which is 3 times larger than that of C{\sc iv}.
If this tentatively detected O{\sc iii]} traces the systematic redshift of J1423$-$0018, the Ly$\alpha$ velocity offset is still small compared to $z>6$ galaxies, which often show $\ll-100$ km s$^{-1}$  \citep[see the compilations of][]{Hashimoto17, Hutchison19}.
More robust detection of O{\sc iii]} or other lines such as  rest-frame far-IR [C{\sc ii}] 158 $\mu$m or [O{\sc iii}] 88 $\mu$m is required to confidentially investigate the Ly$\alpha$ offset of J1423$-$0018.
The flux ratio is O{\sc iii]} $\lambda1661$ / O{\sc iii]} $\lambda1666$ $=1.0$, while \citet{Hainline11} report flux ratios of $3.4$ for their $2<z<3$ type-II objects.
The rest-frame equivalent widths are EW$_\mathrm{rest}=5.3_{-1.5}^{+1.7}$ \AA\ for O{\sc iii]} $\lambda1661$ and $5.3_{-1.4}^{+1.6}$ \AA\ for O{\sc iii]} $\lambda1666$, for a total of $10.6_{-2.1}^{+2.8}$ \AA.
We emphasize again that these O{\sc iii]} measurements must be interpreted with caution.

No significant excess was identified from the other emission lines we covered with MOSFIRE, thus we report  $2\sigma$ upper limits of the fluxes and equivalent widths in Table~\ref{tab:emission_line}.
Those upper limits were derived by integrating continuum-subtracted fluxes of the mock spectra over the wavelength ranges of $\pm 2$ FWHM of C{\sc iv} from the line centers expected from the C{\sc iv} redshift, and measuring the 16$^{\rm th}$ and 84$^{\rm th}$ percentiles.
We could not derive any useful constraints on N{\sc iv]}$\lambda1483$, He{\sc ii}, or Si{\sc iii]}$\lambda 1883$, as those lines overlap with strong sky emission lines.

\section{Discussion} \label{sec:discussion}
The line widths of Ly$\alpha$ (FWHM $<230$ km s$^{-1}$) and C{\sc iv} (FWHM $=120^{+20}_{-20}$ km s$^{-1}$) in J1423$-$0018 are much narrower than those of typical type-I quasars, and even narrower than those of local narrow-line Seyfert 1 galaxies  \citep[FWHM $\sim$ 500--2000 km s$^{-1}$; e.g.,][]{Constantin03}, whereas local Seyfert 2 galaxies \citep{Veilleux91} and low-redshift narrow-line AGNs \citep{Zakamska03, Hao05} show lines with widths $\lesssim500$ km s$^{-1}$.
Therefore, J1423$-$0018 is a good candidate to be a reionization-era analog to classical type-II AGNs.
In this section, we discuss how J1423$-$0018 is compared with various types of type-II AGNs and galaxies in the literature based on the rest-frame UV properties.

\subsection{Ly$\alpha$ and C{\sc iv} properties} \label{sec:dis_lya}
\begin{figure}[tbp]
\centering
 \includegraphics[width=\linewidth]{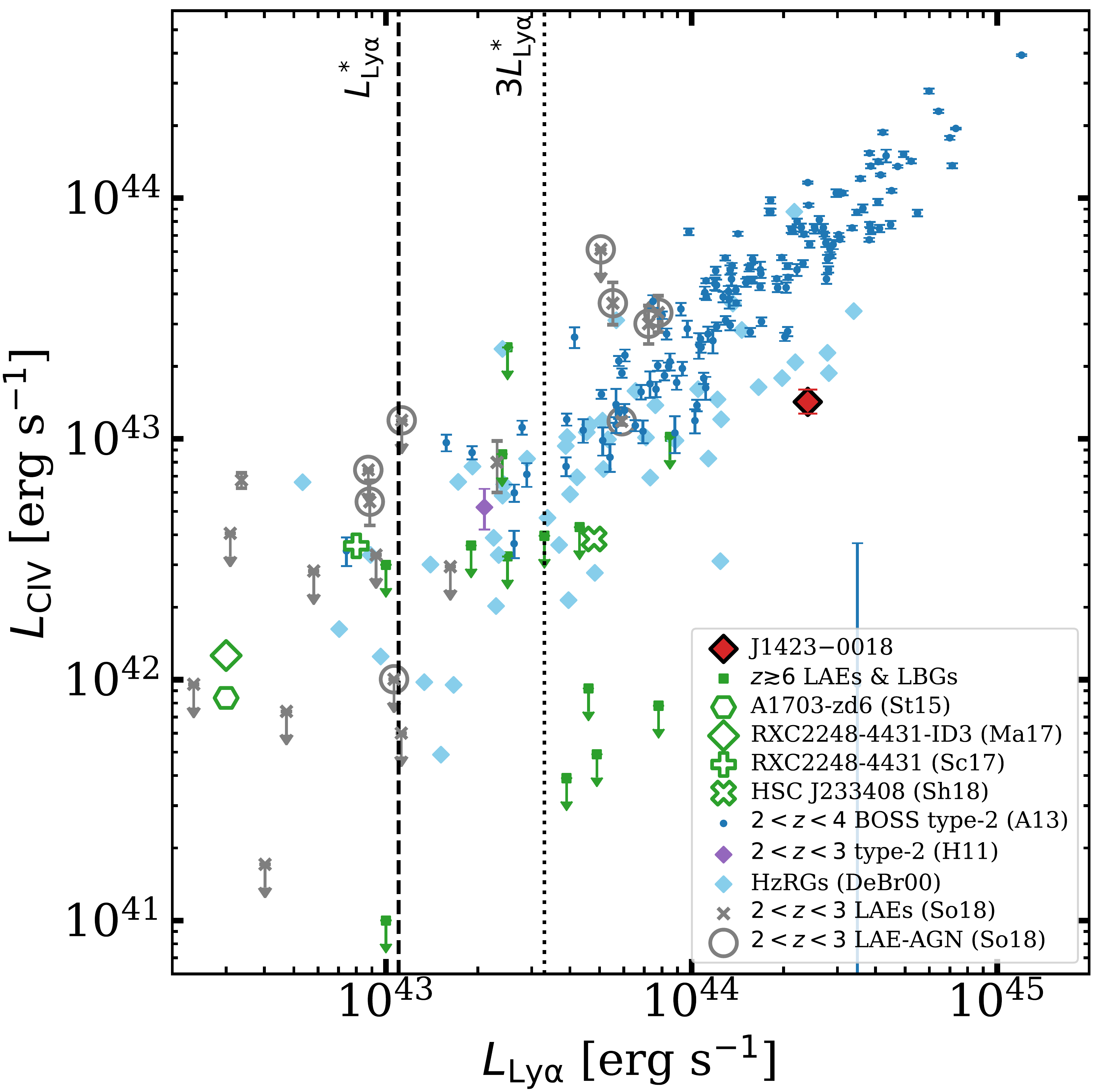}
\caption{
The Ly$\alpha$ and C{\sc iv} luminosities of selected type-II AGNs/quasars and galaxies.
J1423$-$0018 (red diamond) is more luminous than the $z\gtrsim6$ LAEs and LBGs compiled in \citet[][green squares]{Matthee17} in both emission lines, while it is as luminous as the type-II quasar candidates at $2<z<4$ \citep[][blue dots]{Alexandroff13}.
C{\sc iv}-emitting galaxies \citep{Stark15, Mainali17, Schmidt17, Shibuya18} are shown as distinct symbols.
Note that the quoted C{\sc iv} line luminosity of RXC2248-4431-ID3 \citep{Mainali17} is that of the C{\sc iv} $\lambda 1550$ line alone.
The grey crosses are luminous LAEs at $2<z<3$ from \citet{Sobral18}.
Those objects spectroscopically confirmed as AGN are indicated with large grey circles.
Objects from the type-II AGN sample of \citet{Hainline11} at $2<z<3$ are shown in purple.
The sample of high-$z$ radio galaxies (HzRGs) at $0<z<5.2$ from \citet{DeBreuck00} are shown in light blue diamonds, where we show only those object for which they provide both  Ly$\alpha$ and C{\sc iv} line fluxes  (their Table A1). 
The dashed line shows the characteristic Ly$\alpha$ luminosity  $L^*_\mathrm{Ly\alpha}$ at $z=5.7$ \citep[$L^*_\mathrm{Ly\alpha}=1.1\times 10^{43}$ erg s$^{-1}$;][]{Konno18}.
The dotted line shows $3L^*_\mathrm{Ly\alpha}$, above which the Ly$\alpha$ luminosity function has a bright-end excess \citep{Konno18}.
 } \label{fig:fig5}
\end{figure}

We first explore the possibility that J1423$-$0018 is an extreme LAE powered solely by star formation.  In this context, the 
 high luminosity of Ly$\alpha$ ($10^{44.30\pm0.01}$ erg s$^{-1}$) and C{\sc iv} ($10^{43.2\pm0.2}$ erg s$^{-1}$) in J1423$-$0018 are unique.
Figure~\ref{fig:fig5} shows that the Ly$\alpha$ luminosity of J1423$-$0018 is $\gtrsim0.5$ dex higher than objects in the sample of $z=6$--$7$ Lyman break galaxies and LAEs compiled in \citet{Matthee17}.
\citet{Konno18} found that the $z=5.7$ and $z=6.6$ Ly$\alpha$ luminosity functions have an excess relative to their Schechter function fit above $L_\mathrm{Ly\alpha}\sim10^{43.5}$ erg s$^{-1}$.
This bright-end excess could be due to significant AGN contribution, as has been argued at lower redshift \citep{Konno16, Spinoso20, Zhang21}.
Spectroscopic follow-up observations of $2<z<3$ LAEs by \citet{Sobral18} found that the AGN fraction
reaches 100\%\ at $L_\mathrm{Ly\alpha}\gtrsim 2L_\mathrm{Ly\alpha}^*$ and $L_\mathrm{UV}\gtrsim 2L_\mathrm{UV}^*$, where $L_\mathrm{Ly\alpha}^*$ and $L_\mathrm{UV}^*$ are the knee luminosities of the Ly$\alpha$ and the UV luminosity functions, respectively. 
Their AGN/galaxy classification was based on strong N{\sc v} emission (N{\sc v}/Ly$\alpha$ $>0.1$), and line ratio diagnostics based on C{\sc iv}/He{\sc ii} and C{\sc iii]}/He{\sc ii}.
J1423$-$0018 satisfies both of the $L_\mathrm{Ly\alpha}$ and $L_\mathrm{UV}$ thresholds at $z\sim6$, with $L_\mathrm{Ly\alpha}=18 L^*_\mathrm{Ly\alpha}$ \citep[$L_\mathrm{Ly\alpha}^*=1.1\times10^{43}$ erg s$^{-1}$ at $z=5.7$;][]{Konno18}, and $L_\mathrm{UV}=2.5 L^*_\mathrm{UV}$ \citep[$M_\mathrm{UV}^{*}=-20.9$ mag;][]{Ono18}.
\citet{Calhau20} used X-ray and radio detections to show that the AGN fraction of LAEs decreases toward high redshift. 
Although their constraints are poor at $z\sim3.5$--$6$, objects in their sample with luminosity above $L_\mathrm{Ly\alpha}=10^{44}$ erg s$^{-1}$ and $z\sim2$--$6$ are dominated by AGN.  
However, the high luminosities of the UV continuum and Ly$\alpha$ alone may not be sufficient to prove that J1423$-$0018 is an AGN.  
For example, \citet{MC20} discovered a blue star-forming galaxy at $z=2$ (BOSS-EUVLG1) that has $L_\mathrm{Ly\alpha}=10^{44.0\pm0.1}$ erg s$^{-1}$, comparable to J1423$-$0018 (See Figs.~\ref{fig:fig1} and \ref{fig:fig2}).
It is unlikely that this particular source is the same population as J1423$-$0018, because the continuum luminosity of BOSS-EUVLG1 ($M_\mathrm{UV}=-24.40\pm0.05$ mag) is considerably higher than that of J1423$-$0018, rather close to broad-line quasars and the BOSS type-II AGN candidates \citep{Alexandroff13}.

After Ly$\alpha$, C{\sc iv} is usually the second brightest emission line in the rest-UV spectrum of  AGN.  
While C{\sc iv} has a high ionization potential (47.9 eV), it can arise from hard ionizing radiation of either an AGN or a young metal-poor galaxy.
Such C{\sc iv}-luminous star-forming galaxies should be common in the reionization era, and indeed 
C{\sc iv} has recently been detected in galaxy samples at $z>6$: A1703-zd6 \citep{Stark15}, RXC2246-4431 \citep{Schmidt17}, RXC2246-4431-ID3 \citep{Mainali17}, and HSC J233408 \citep{Shibuya18}\footnote{The C{\sc iv}$\lambda1550$ detection of HSC J233408 is tentative.}.
The absolute UV magnitudes of those C{\sc iv} emitters are $\gtrsim2$ mag fainter than that of J1423$-$0018.
Figure~\ref{fig:fig5} shows that the C{\sc iv} luminosity of J1423$-$0018 is considerably higher than C{\sc iv}-emitting $z=6$--$7$ galaxies, and is comparable to those of the $2<z<4$ type-II quasars of \citet{Alexandroff13}, the high-redshift ($0<z<5.2$) radio galaxies (HzRGs) of \citet{DeBreuck00}, and the $2<z<3$ LAE-AGNs of \citet{Sobral18}.

\citet{Nakajima18} show that star-forming galaxies can power C{\sc iv} up to EW$_\mathrm{rest}=12$ \AA\ within plausible parameter ranges of metallicity and ionization hardness.  
Indeed, the C{\sc iv} emission from known intermediate-redshift type-II AGNs/quasars mostly lies above the threshold  \citep{DeBreuck00, Hainline11, Alexandroff13, LeFevre19, Mignoli19}.
Thus the observed EW(C{\sc iv})$_\mathrm{rest}=37^{+6}_{-5}$ \AA\ of J1423$-$0018 strongly suggests that the C{\sc iv}-emitting gas is powered by an AGN rather than star formation.

\begin{figure*}[htbp]
\centering
 \includegraphics[width=\linewidth]{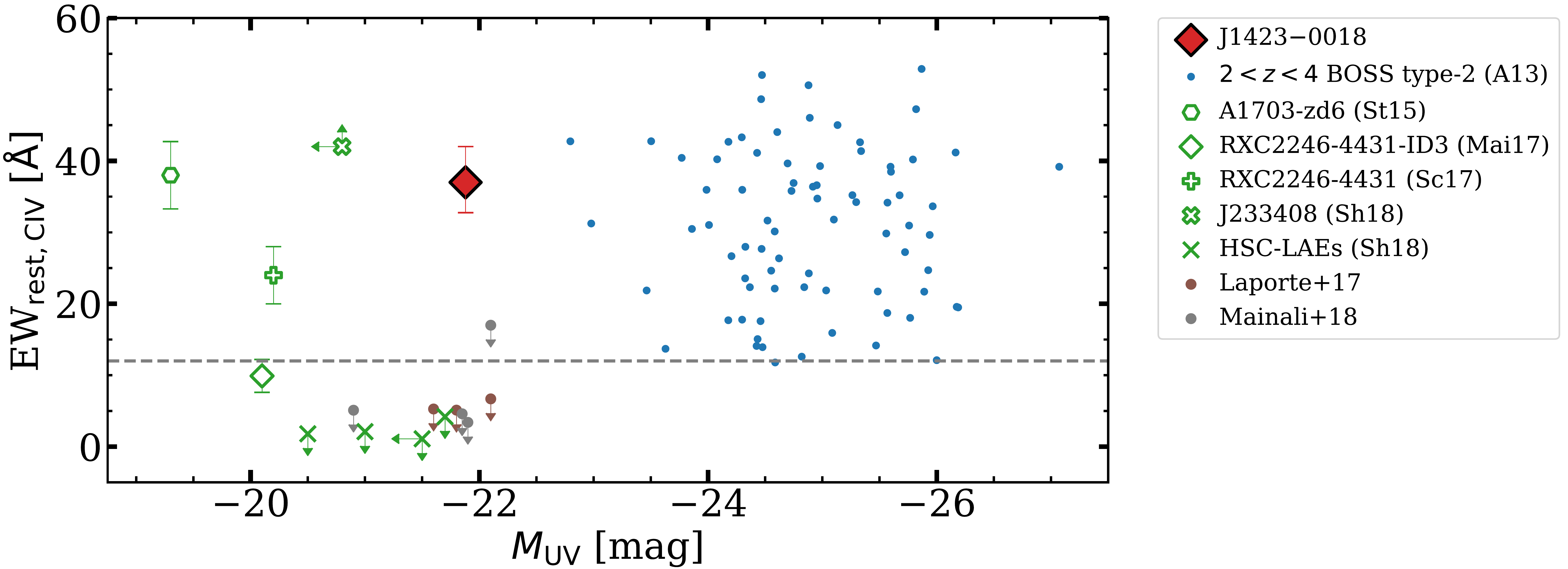}
\caption{C{\sc iv} rest-frame EW as a function of UV continuum absolute magnitude.
The symbols are the same as Figure~\ref{fig:fig5}, while we added  $z\sim6$--$7$ galaxies from \citet[][brown]{Laporte17} and \citet[][grey]{Mainali18}.
The horizontal dashed line is drawn at EW $=12$ \AA, the upper limit for star-formation powered emission \citep{Nakajima18}.
} \label{fig:fig6}
\end{figure*}

Figure~\ref{fig:fig6} shows C{\sc iv} rest-frame EW as a function of UV magnitude for various high-redshift populations.  
This figure shows that some of the strongest C{\sc iv} emitters known at $z \sim6$--$7$ \citep{Stark15, Schmidt17, Mainali17, Shibuya18} reside in the faintest UV magnitude range ($M_\mathrm{UV}\gtrsim-20$ mag).
This may reflect the fact that metal-poor galaxies are often faint in UV and low-mass, while galaxies with C{\sc iv}  EW$_\mathrm{rest}(\mathrm{ C{\sc IV}})\geq 12$ \AA\  are at least partly powered by AGNs in the framework of \citet{Nakajima18}.
Although J1423$-$0018 has a C{\sc iv} EW considerably higher than most sources at similar UV magnitudes \citep{Laporte17, Schmidt17, Shibuya18, Mainali18}, it is  comparable to those of the BOSS type-II quasar candidates with EW$_\mathrm{rest}(\mathrm{ C{\sc IV}})\approx12$--$50$ \AA\  \citep{Alexandroff13}.

\begin{figure*}[htbp]
\centering
 \includegraphics[width=\linewidth]{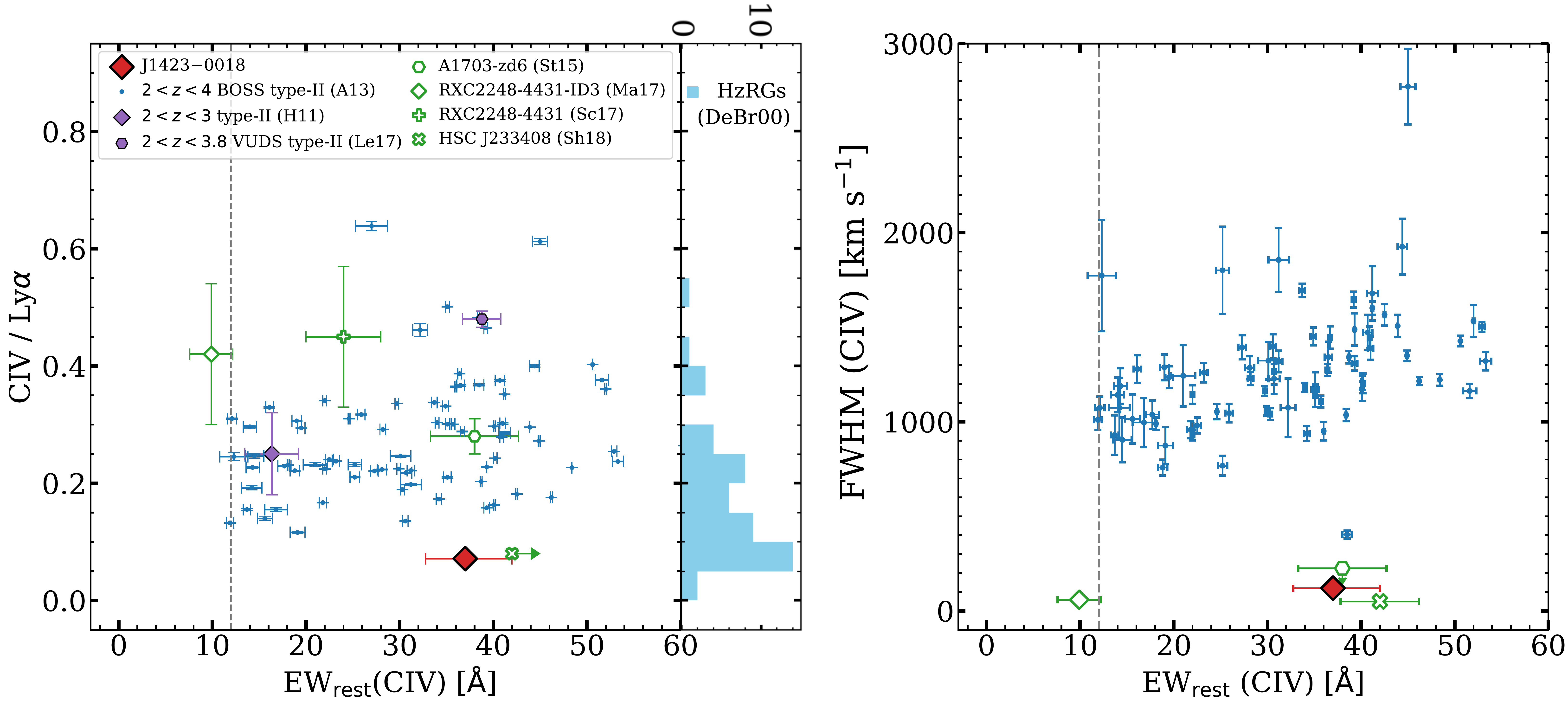}
 \caption{
 ({\it a}) 
 The relationship between the rest-frame C{\sc iv} equivalent widths and the C{\sc iv}/Ly$\alpha$ flux ratios.
 The same samples as in Figure~\ref{fig:fig5} are shown. Galaxies detected in C{\sc iv} are plotted in green \citep{Stark15, Mainali17, Schmidt17, Shibuya18}.
 We also added the data point of the composite spectrum of $2<z<3.8$ narrow-line AGNs in the VUDS field as a purple hexagon \citep{LeFevre19}.
 The right-hand histogram shows the $0<z<5.2$ HzRGs of \citet{DeBreuck00}.  
 ({\it b})
The FWHM of C{\sc iv} versus its rest-frame equivalent width.
The vertical dashed line at EW$_\mathrm{rest}$(C{\sc iv})$=12$ \AA\ indicates the boundary above which only AGNs can fall, according to the photoionization model of \citet{Nakajima18}.
}\label{fig:fig7}
\end{figure*}

It is not clear whether J1423$-$0018 is a member of the same classes of type-II AGNs known at lower redshift, because the nature of type-II AGNs is highly dependent on the method by which they were selected.
Figure~\ref{fig:fig7} compares the C{\sc iv} rest-frame EW, the flux ratio of C{\sc iv} to Ly$\alpha$, and C{\sc iv} line width of J1423$-$0018 with those of intermediate-redshift ($z\sim2$--$5$) type-II AGNs \citep{DeBreuck00,Hainline11, Alexandroff13, LeFevre19}.
The ratio of C{\sc iv} to Ly$\alpha$ ($0.08\pm0.01$) in J1423$-$0018 is significantly smaller than those optically-selected type-II AGNs, which lie in the range $\approx0.1$--$0.6$; even though  intergalactic medium absorption at Ly$\alpha$ is stronger at higher redshift.

We should note that the objects in the \citet{Alexandroff13} sample of intermediate-redshift Type-II candidates is likely to be only modestly obscured.  
These objects show higher UV luminosity, their C{\sc iv} line widths are FWHM $=1000$--$2000$ km s$^{-1}$, approaching values for broad-line quasars, and NIR spectroscopy shows broad H$\alpha$ in many cases \citep[up to 7500 km s$^{-1}$;][]{Greene14}. 
The narrow-line AGNs of \citet{Hainline11}, \citet{Mignoli19}, and \citet{LeFevre19} are likely to be more similar  to J1423$-$0018: their UV continuum luminosities are comparable to the SHELLQs narrow-line population, and their composite spectra do not apparently show broad components in emission lines (FWHM $<1350$ km s$^{-1}$; \citealt{Mignoli19}).

\subsection{Similarity to High-redshift Radio Galaxies} \label{sec:dis_HzRG}
High-redshift radio galaxies (HzRGs), one known  population of type-II AGNs, tend to have relatively small C{\sc iv}/Ly$\alpha$ ratios \citep[e.g.,][]{DeBreuck00, VM07, MatsuokaK09, MatsuokaK11}.
\citet{DeBreuck00} compiled rest-frame UV spectra of a large sample of spectroscopicaly confirmed  HzRGs at $0<z<5.2$.  
They found a median C{\sc iv}/Ly$\alpha$ ratio of about 0.1, with some objects as low as 0.03. 
There are objects in the HzRG sample with Ly$\alpha$ rest-frame EWs up to $\approx1000$ \AA, even higher  than that of J1423$-$0018. 
Thus, the emission-line properties of J1423$-$0018 are similar to those of HzRGs.

We searched for the radio counterpart of J1423$-$0018 in the 1.4 GHz radio source catalog of the Faint Images of the Radio Sky at Twenty-cm survey (FIRST, \citealt{FIRST}).
There is no source entry within 30 arcseconds of the optical position of J1423$-$0018, which corresponds to a flux upper limit of $1$ mJy.
The HzRG sample of \citet{DeBreuck00a}, one of the HzRG samples compiled by \citet{DeBreuck00}, 
applied a 1.4 GHz flux limit of $>10$ mJy to select HzRG candidates with steep radio slopes between 1.4 GHz and 325 MHz; therefore, J1423$-$0018 does not meet this criteria.
The only known $z\sim6$ radio galaxy, TGSS1530 \citep[][$z=5.7$]{Saxena18} has a 1.4 GHz flux of $7.5\pm0.1$ mJy in the FIRST source catalog. 
The Ly$\alpha$ luminosity of TGSS 1530 is $L_\mathrm{Ly\alpha}=5.7\pm0.7\times10^{42}$ erg s$^{-1}$, while its C{\sc iv} strength has not yet been reported.  
This Ly$\alpha$ luminosity is smaller than that of J1423$-$0018 by a factor of 35.

In the right panel of Figure~\ref{fig:fig7}a we show the distribution of C{\sc iv}/Ly$\alpha$ for the HzRG sample of \citet{DeBreuck00}.
The rest-frame C{\sc iv} EWs of this sample span a wide range, from 0 to $\approx 100$ \AA.
We do not show their UV magnitudes and C{\sc iv} EWs in Figure~\ref{fig:fig6} and Figure~\ref{fig:fig7}a because  those properties are not provided for the objects in the sample.
\citet{DeBreuck00} shows that the C{\sc iv}/Ly$\alpha$ ratios of their $z \gtrsim 3$ sample are about a half of those seen in their lower-redshift sample.  
This may reflect a lower ionized gas metallicity  at high redshift \citep{DeBreuck00}, or host star formation responsible for a larger fraction of the strong Ly$\alpha$ emission at higher redshift \citep[][ see Section~\ref{sec:conclusion}]{VM07}.
We note that there is a Ly$\alpha$-only AGN known at $z=2$ \citep{Hall04}, although its Ly$\alpha$ is broad (FWHM $=1400$ km s$^{-1}$), and its EW(Ly$\alpha$) ($=34$ \AA) is not as extreme as J1423$-$0018.

\subsection{C{\sc iv} Emitters at $z=6$--$7$} \label{sec:dis_CIV}
Figure~\ref{fig:fig7}a also shows that $z=6$--$7$ C{\sc iv}-emitting objects from galaxy samples share common properties with  intermediate-$z$ type-II AGNs and J1423$-$0018.
The C{\sc iv} emitters with rest-frame EW $\sim20$ \AA\ or larger are classified as AGNs according to the diagnostics of \citet{Nakajima18}\footnote{The object RXC2246-4431-ID3 has  EW(C{\sc iv}$\lambda1550$)$=9.9\pm2.3$ \AA\ \citep{Mainali17}.  The $\lambda1548$ line is affected by a strong sky line, and is thus not measured.
\citet{Mainali17} carried out emission line diagnostics based on C{\sc iv}/He{\sc ii} and O{\sc iii]}/He{\sc ii} to suggest that this particular object is primarily powered by star formation, using the photoionization models of \citet{Feltre16}.}.
\citet{Stark15} argue that the only way to model the photoionization of  A1703-zd6, a C{\sc iv} emitter at $z=7.043$ with EW(C{\sc iv}) $\sim40$ \AA\ with stellar ionization is to assume a  low metallicity ($12+\log{\mathrm{O/H}}=7.0$) and a high ionization parameter ($\log{U}=-1.4$).  Pure AGN models can also explain its observed O{\sc iii]}$/$C{\sc iv} and He{\sc ii}$/$C{\sc iv} line ratios.
\citet{Nakajima18} pointed out that the relative abundance ratio of C/O needs to be near solar or super-solar to realize the observed C{\sc iv} EW using stellar radiation alone.  
This would be surprising, given their standard models of metal-poor galaxies.
Given this controversial situation, detection of other high-ionization metal emission lines such as He{\sc ii} and C{\sc iii]}
would provide firmer evidence whether J1423$-$0018 and those C{\sc iv} emitters are the same populations of AGNs or galaxies.

If J1423$-$0018 is a galaxy powered principally by star formation, the high EW(Ly$\alpha$) $=370\pm30$ \AA\ (which could be twice as large after correcting for the strong IGM absorption at $z=6.1$) suggests that the stellar population should be young ($\ll10^7$ yr) and metal-poor ($Z\ll0.02Z_\odot$) according to the burst star-formation model of \citet{Hashimoto17}.  

There are several other extreme Ly$\alpha$ emitting galaxies known at $z>5$ \citep{Kashikawa12, McGreer18} with rest-frame Ly$\alpha$ EWs of $>200$ \AA.  
However, those authors noted that C{\sc iv} is not detected in these sources, and concluded that they are not powered by AGNs. 
At this stage, the relation between those emission line galaxies and J1423$-$0018 is unclear.

\begin{figure}[tbp]
\centering
 \includegraphics[width=\linewidth]{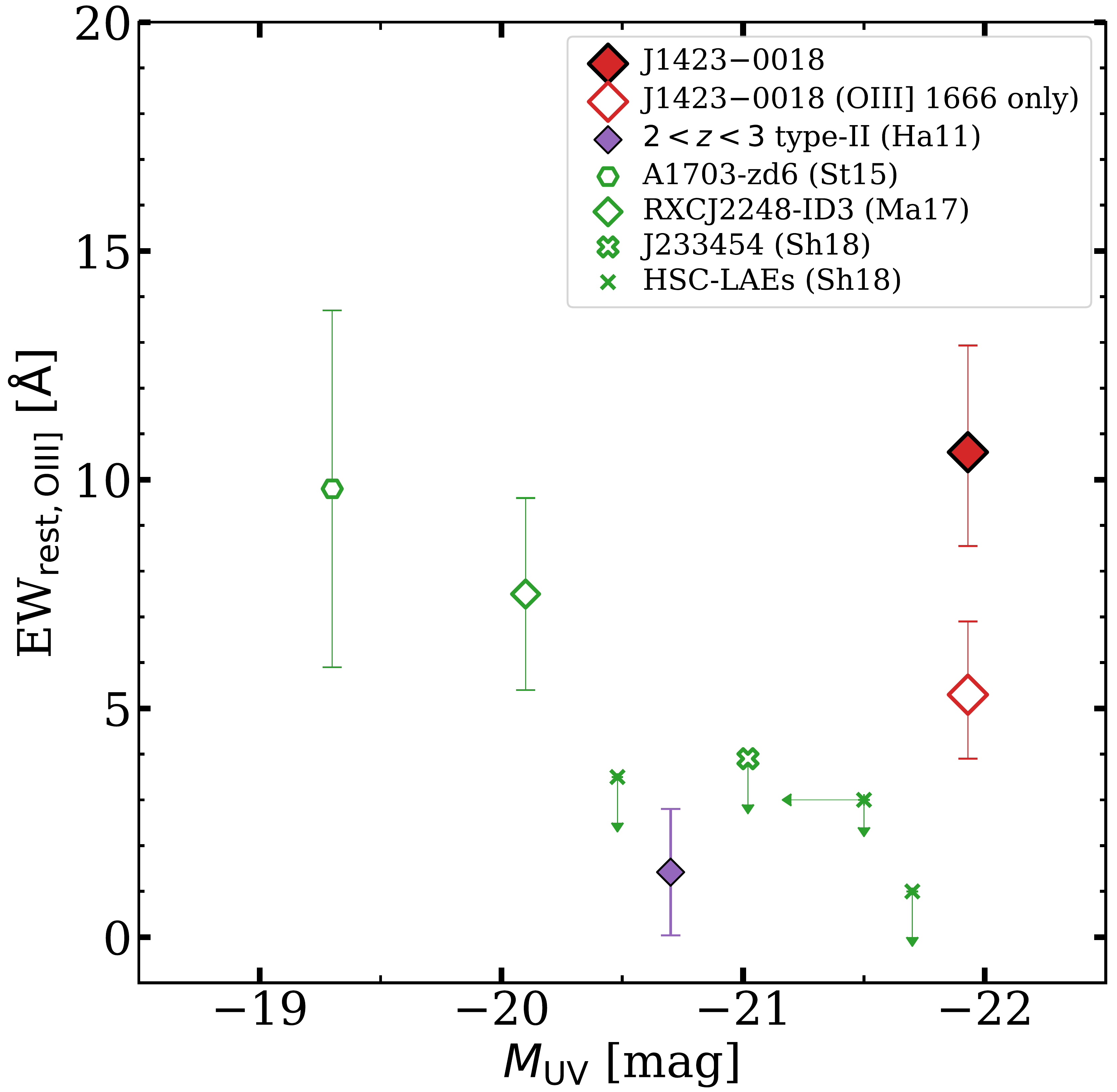}
\caption{ O{\sc iii]} rest-frame EW of various galaxy samples as a function of UV magnitude.
For J1423$-$0018 we show two cases.
The filled diamond shows the case where we consider both O{\sc iii]} $\lambda1661$ and O{\sc iii]} $\lambda1666$, while the open diamond shows the case where we consider O{\sc iii]} $\lambda1661$.
The other symbols are the same as Figure~\ref{fig:fig6}.
} \label{fig:fig8}
\end{figure}

\subsection{O{\sc iii]} Properties} \label{sec:dis_OIII}
Figure~\ref{fig:fig8} shows the rest-frame O{\sc iii]} EWs  as a function of UV magnitude for the samples of low-redshift type-II quasars and the $z\sim6$--$7$ galaxies we have considered in Figure~\ref{fig:fig6}.  
The figure includes the tentative detection of  O{\sc iii]} $\lambda\lambda1661,1666$ in J1423$-$0018.
Given the spatial offset of O{\sc iii]} $\lambda1661$ (Section~\ref{sec:obs} and Figure~\ref{fig:fig3}), we consider two cases: one in which we assume that the  O{\sc iii]} $\lambda1661$ detection is real, and the other in which we only consider 
O{\sc iii]} $\lambda1666$.
Just as we found for C{\sc iv}, the observed total rest-frame EW ($=10.6$ \AA\ or $=5.3$ \AA) is as strong as the UV-fainter C{\sc iv} emitters of \citet{Stark15} and \citet{Mainali17} at $z\sim6$--$7$, whereas it is a factor of seven (or five) larger than that of the \citet{Hainline11} type-II AGNs at $2<z<3$.
The strong O{\sc iii]} of J1423$-$0018, if real, may suggest a distinct change in the hardness of the ionizing radiation of type-II AGNs at high redshift; however, the ionization parameter of J1423$-$0018 is hard to constrain from the currently available data, especially due to the lack of high-ionization lines such as He{\sc ii} and N{\sc v}.

We compare the O{\sc iii]}/C{\sc iv} flux ratio of J1423$-$0018 ($=0.24$ or $0.12$) to other galaxy and type-II AGN samples.
Local metal-poor galaxies \citep{Berg19a} show relatively strong O{\sc iii]}, with O{\sc iii]}/C{\sc iv} line ratios of $0.5$--$1.9$.
The $z\sim6$--$7$ C{\sc iv}-detected Lyman break galaxies of \citet{Stark15} and  \citet{Mainali17} have comparable O{\sc iii]}/C{\sc iv} to this sample.
On the other hands,  the intermediate-redshift type-II AGNs of \citet{Hainline11} and \citet{Alexandroff13} show a dex fainter O{\sc iii]}/C{\sc iv} (see Figure~4 of \citealt{Mainali17}).
J1423$-$0018 lies in between the metal-poor star-forming galaxies and low-redshift type-II AGNs.

\section{Conclusion and Future Prospects}\label{sec:conclusion}
The Ly$\alpha$ and C{\sc iv} emission line properties (line width, luminosity and EW) of J1423$-$0018 suggest that this object is an obscured type-II AGN at $z=6.1292$.  However, further observations are needed to confirm the type-II nature and, if it is an obscured AGN, determine how extincted it is, its bolometric luminosity, and the contribution of the host galaxy to the observed rest-UV continuum and emission lines.
Photoionization models suggest that measurements of equivalent widths and line ratios of additional lines (including  C{\sc iii]}, He{\sc ii}, and N{\sc v}) would lead to more definitive conclusions \citep[e.g.,][]{Feltre16, Byler18, Nakajima18}. 
We were not able to observe these lines given our spectral coverage and contamination from strong sky lines.  
It is thus important to observe analogs to J1423$-$0018 at different redshifts to constrain their properties. 
Spectroscopic follow-up observations with {\it the James Webb Space Telescope} (JWST) would enable us to carry out the classic AGN  diagnostic test based on the rest-frame optical emission lines.
We note that J114658.89$-$000537.7, one of the five extremely Ly$\alpha$ luminous narrow-line objects among the SHELLQs sample (Figure~\ref{fig:fig1}a, \ref{fig:fig2}) will be observed in an approved Cycle 1 General Observers program by \citet{Onoue21_JWST} as a part of their 12 targets.
X-ray detection would provide the most direct signature of being an AGN, and if sensitive enough, we could also estimate the hydrogen column density.
We are planning to obtain deep X-ray data of the five extreme narrow-line objects (including J1423$-$0018) with {\em Chandra}.
Deep radio follow-up observations will test if J1423$-$0018 has a significant synchrotron radiation as those of HzRGs.

Finally, it is possible that the observed rest-UV continuum and emission lines of the SHELLQs narrow-line objects are powered by {\it both} AGN and star formation.
This scenario is suggested by the small C{\sc iv}/Ly$\alpha$ line ratio of J1423-0018, because \citet{VM07} argued that the strong Ly$\alpha$ emission of HzRGs is partly contributed by gas irradiated by the young stellar population common in high-redshift galaxies.
Such an intermediate type of high-redshift source would naturally appear in the UV magnitude range where galaxies are more common than AGNs \citep{Adams20, Bowler21}.
The cosmological simulations of \citet{Trebitsch20} support this hypothesis, which showed that such AGN/galaxy composite objects are commonly found in UV bright galaxies at $z\gtrsim6$.
Since the attenuation of nuclear emission to the UV wavelength ranges is not constrained for J1423$-$0018, a higher signal-to-noise ratio NIR spectrum and multi-wavelength photometry of the narrow-line objects are required to disentangle the observed continuum and emission lines from AGNs and host galaxies.
Far-infrared follow-up observations will provide clues to the host star-formation activities.
Such observations would allow us to diagnose whether the obscuration of the nucleus is due to dust at the core of the galaxy, or distributed throughout the host.

\acknowledgments

The data presented herein were obtained at the W. M. Keck Observatory, which is operated as a scientific partnership among the California Institute of Technology, the University of California and the National Aeronautics and Space Administration. The Observatory was made possible by the generous financial support of the W. M. Keck Foundation.
We wish to recognize and acknowledge the very significant cultural role and reverence that the summit of Maunakea has always had within the indigenous Hawaiian community.  We are most fortunate to have the opportunity to conduct observations from this mountain.

This publication has made use of data from the VIKING survey from VISTA at the ESO Paranal Observatory, programme ID 179.A-2004. 
Data processing has been contributed by the VISTA Data Flow System at CASU, Cambridge and WFAU, Edinburgh.

The Hyper Suprime-Cam (HSC) collaboration includes the astronomical
communities of Japan and Taiwan, and Princeton University.  The HSC
instrumentation and software were developed by the National
Astronomical Observatory of Japan (NAOJ), the Kavli Institute for the
Physics and Mathematics of the Universe (Kavli IPMU), the University
of Tokyo, the High Energy Accelerator Research Organization (KEK), the
Academia Sinica Institute for Astronomy and Astrophysics in Taiwan
(ASIAA), and Princeton University.  Funding was contributed by the FIRST 
program from Japanese Cabinet Office, the Ministry of Education, Culture, 
Sports, Science and Technology (MEXT), the Japan Society for the 
Promotion of Science (JSPS),  Japan Science and Technology Agency 
(JST),  the Toray Science  Foundation, NAOJ, Kavli IPMU, KEK, ASIAA,  
and Princeton University.

The authors wish to thank Ramesh Mainali for useful comments on RXCJ2248-ID3.
We also thank the anonymous referee for providing us constructive comments on the draft.

This work was supported by JSPS KAKENHI Grant Numbers 17H04830 (YM), 20H01949 (TN), 18J01050 (YT).
MO was supported by the ERC Advanced Grant 740246 “Cosmic gas.”
YM was supported  by the Mitsubishi Foundation Grant No. 30140.
TH was supported by Leading Initiative for Excellent Young Researchers, MEXT, Japan (HJH02007).

\vspace{5mm}
\facilities{Keck:I (MOSFIRE)}

\software{Astropy \citep{Astropy, Astropy18}, Matplotlib \citep{matplotlib}, NumPy \citep{Numpy20}, SciPy \citep{Scipy20}
    }

\end{document}